\documentclass[12pt,a4paper ]{article}
\usepackage[cp866]{inputenc}

\usepackage{graphicx}

\input epsf

\newcommand{\frat}[2]{\frac{\textstyle #1}{\textstyle #2}}
\newcommand{\vf}[1]{\mbox{\boldmath $#1$}}

\begin{document}

\begin{center}
{\Large \bf Exploring meson correlators in the model with four-quark interaction
Lagrangian}\\
 \vspace{0.5cm} S. V. Molodtsov$^{1,2}$, M. K. Volkov$^1$, G. M. Zinovjev$^3$
\\
\vspace{0.5cm}
{\small $^1$Joint Institute for Nuclear Research, Dubna, Moscow region,
RUSSIA}\\
\vspace{0.5cm}
{\small $^2$Institute of Theoretical and Experimental Physics, Moscow, RUSSIA}\\
\vspace{0.5cm}
{\small $^3$Bogolyubov Institute for Theoretical Physics, National
Academy of Sciences of Ukraine, Kiev, UKRAINE}
\end{center}
\vspace{0.5cm}

\begin{center}
\begin{tabular}{p{16cm}}
{\small{Meson correlation functions are studied in the model with four-fermion
interaction Lagrangian. We demonstrate that despite the singular character of
system mean energy and corresponding quark condensate found, the meson
observables are finite, quite well identified and
compatible with experimental energy scale. It allows the similar model
Hamiltonians to be used for describing the nonequilibrium features of
quark/hadron systems which reveal themselves in studying ultrarelativistic
heavy ion collisions. The analytical results for meson correlation
functions in the Keldysh model are given.
}}
\end{tabular}
\end{center}
\vspace{0.5cm}

The intensive experimental study of ultrarelativistic heavy ion collisions
pushes forward a great interest in developing theoretical and phenomenological
description of nonequilibrium processes in quark/hadron matter. This stage
preceding a thermalization and chemical equlibration is of crucial
importance in governing the collision process development but very complicated
for reliable theoretical interpretation. The analysis of experimental data
available
leads to the conclusion that at the initial moment
of clashing, system with great number of degrees of freedom appears and its
constituents are strongly interacting. The characteristic time of nonequilibrium
stage at the RHIC
experiments is roughly estimated, for example, as $1$ fm/$c$ \cite{RHIC} and the
energy density reached exceeds $15$ GeV/fm$^3$ which is much higher than the
corresponding quantity for the
nuclear matter and its value for the bag model. Clearly, such estimates urge
(and allow)
to speculate on the description of processes at the quark level with the
mechanism of dynamical
mass generation included.

In such a context the study of equilibrated states and phase diagram of
quark/hadron matter is based on the Nambu--Jona-Lasinio model (NJL) \cite{nj}
in which the adequate picture of spontaneous chiral symmetry breaking is
properly incorporated and low energy meson physics is well understood \cite{vmk},
\cite{NJL}. However, the NJL model does not accommodate the gluon degrees
of freedom and is not directly applicable for analyzing the
nonequilibrium processes. Thus, searching the related models which are free of
such shortages and share the attractive features of NJL in the low energy
region is a topical and practical task.
In Ref. \cite{MZ} the effective Hamiltonians with four-fermion interaction in
the form of the product of two spatially separated currents mediated by
a formfactor have been considered. As in the NJL model it is supposed
that the ground state of the system is formed by the quark--anti-quark pairs
with vacuum quantum numbers and oppositely directed momenta and in the framework
of the Bogolyubov--Hartree--Fock approximation the description of quarks
as the quasi-particles which is appropriate in the broad momentum range
of momenta has been developed. A comparative analysis of the models with the
formfactor behaving as the $\delta$-function in the coordinate space
(the NJL model) and the similar formfactor behaviour but in the momentum space
(the model allied to the Keldysh model which is well known in the condensed
matter physics) \cite{k}) teaches that these dissimilar models lead to the
equivalent quasi-particles when the dynamical quark masses are comparable.
Actually, it turns out the parameters characterizing the quasiparticles are
developed by the common dynamical mechanism which is practically insensitive
to the formfactor species. One unexpected feature of these models is
a discontinuity of mean energy functional considered as a function of current
quark mass what results in some difficulties at fitting the quark condensate
beyond the chiral limit. Speaking literally the quark condensate
and the mean energy of quark ensemble are infinite. Fortunately, neither are
physically observable quantities and in order to make a reliable
conclusion about the models we should study the meson
correlation functions, for example. In this note we explore the meson
observables in the Keldysh model which resembles a toy model but in some
aspects elucidated below, turns out quite instructive.
The meson characteristics despite the singular character of the ground state are
finite and fairly adequate to correspond to the energy scale of existing
experimental data. Certainly, such a conclusion does not concern all the
features which is quite understandable within so a simple model. For example,
the
$\pi$-meson mass is slightly underestimated when the tuning parameters leading
to the dynamical quark mass
compatible with the NJL result are used and the pion decay constant
$f_\pi$ disappears (see below).

\section{Model Lagrangian}
We take the model Lagrangian density (discussed in Ref. \cite{MZ}) in the form
of a product of two quark currents localized at the space coordinates ${\vf x}$
and ${\vf y}$ which are bound by the formfactor $F_{\mu\nu}$, i.e.
\begin{equation}
\label{sb10}
{\cal L}=\bar q~(i \gamma_\mu \partial_\mu+im)~q-\bar q~t^a\gamma_\mu~q~
\int d{\vf y}~F_{\mu\nu}({\vf x}-{\vf y})~\bar q'~t^a\gamma_\nu~q'~,
\end{equation}
where $q=q({\vf x},t)$, $\bar q=\bar q({\vf x},t)$, $q'=q({\vf y},t)$,
$\bar q'=\bar q({\vf y},t)$ are the (anti-)quark fields,
$t^a=\lambda^a/2$ are the generators of $SU(N_c)$ colour gauge group and $m$ is
a current quark mass.
The Lagrangian density is given in the context of the Euclidean field theory and
$\gamma_\mu$ are the Hermitian Dirac matrices, $\mu,\nu=1,2,3,4$.
The effective Hamiltonian corresponding to the Eq. (\ref{sb10}) results from the
averaging procedure when the quark behaviour is strongly affected by intensive
stochastic gluon field. The (anti-)instanton ensemble was considered as such a
background.
As a general form the formfactor in Eq. (\ref{sb10}) can be presented by a
sum of two components $F_{\mu\nu}({\vf x}-{\vf y})=G~F({\vf x}-{\vf
y})~\delta_{\mu\nu}+J_{\mu\nu}({\vf x}-{\vf y})$,
where the second term is spanned on the relative distance vector.
In the first component we single out the constant $G$ characterizing the
strength  of four-fermion interaction. In particular cases when the formfactor
$F$ has the $\delta$-function shape in coordinate space we come to
the NJL model. With the formfactor behaving as $F({\vf p})=(2\pi)^3~\delta({\vf
p})$
the model is similar to the Keldysh model.
In order to simplify the consideration we ignore the
contribution of the correlator corresponding to $J_{\mu\nu}(p)$.
The unprejudiced analysis of the system behaviour beyond the chiral limit
performed in Ref.\cite{MZ} demonstrates that the four-fermion interaction
develops a singularity. The mean energy of the ensemble goes to
infinity and the quark condensate demonstrates the singular behaviour as well.
In the leading order in the $N_c$-expansion we obtain for the generators of
colour group $\sum^{N_c^{2}-
1}_{a=1}t^a_{ij}t^a_{kl}\approx\frac{1}{2}~\delta_{il}\delta_{kj}$ and utilizing
the Fierz transformation $\gamma_\mu \bigotimes \gamma_\mu=1 \bigotimes 1
+i\gamma_5
\bigotimes i\gamma_5-\frac{1}{2}\gamma_\mu \bigotimes \gamma_\mu -
\frac{1}{2}\gamma_\mu\gamma_5 \bigotimes \gamma_\mu\gamma_5$,
we have for the scalar contribution in the mean field approximation the
following effective Lagrangian density
\begin{equation}
\label{sb11}
{\cal L}\approx\bar q~(i \gamma_\mu \partial_\mu+im)~q-\int d{\vf y}~G~F({\vf
x}-{\vf y})~
\langle \bar q~q'\rangle~\bar q'~q~.
\end{equation}
where the angle brackets denote the corresponding average.
It is interesting to notice the interaction term is composed with the colourless
quark operators with ${\vf x}$ and ${\vf y}$ coordinates interchanged.
The selfconsistency condition allows us to extract the dynamical quark mass as
\begin{equation}
\label{sb13}
M({\vf p})=2N_c~\int \frat{d {\vf q}}{(2\pi)^3}~G~F({\vf p}-{\vf q})~
\frat{m+M({\vf q})}{\left[{\vf q}^2+(m+M({\vf q}))^2\right]^{1/2}}~.
\end{equation}
At the spontaneous breaking of chiral symmetry takes place at this stage we have
for the formfactor behaving as $F({\vf x})=\delta({\vf x})$ the well known gap
equation
$$M=2N_cG~\int^{\Lambda_{{\mbox{\tiny NJL}}}} \frat{d {\vf q}}{(2\pi)^3}
\frat{m+M}{\left[{\vf q}^2+(m+M)^2\right]^{1/2}}~,$$
where $\Lambda_{{\mbox{\tiny NJL}}}$ is the cut off parameter. For the Keldysh
model in the mean field approximation it looks like
\begin{equation}
\label{sb14}
M({\vf p})=2N_cG~\frat{m+M({\vf p})}{\left[{\vf p}^2+(m+M({\vf
p}))^2\right]^{1/2}}~.
\end{equation}
Transforming this solution into the function $p(M)$ we have for the quark mass
in the chiral limit in the Keldysh model $M(p)=\left[(2N_cG)^2-p^2\right]^{1/2}$.
In the further analysis bearing in mind the Fierz transformation we consider a
more
general form of the Lagrangian with two constants included. They
characterize the interacting strength for the scalar and pseudoscalar
channels --- $G$ and for the vector and the axial-vector ones ---
$G_{\mbox{\tiny V}}$.
Clearly, the scalar and pseudo-scalar coupling constats are identical in the
chiral limit \cite{nj}.

\section{Bosonization}
Here we introduce the meson fields adapting well known bosonization procedure
for the scalar channel (similar relations are valid for the other channels).
It is convenient to introduce the auxiliary variables
\begin{eqnarray}
\label{b1}
&&Q_s({\vf x},{\vf y};t)=F^{1/2}({\vf x}-{\vf y})~\frac12~
\left[ ~\bar q({\vf x};t)q({\vf y};t)+\bar q({\vf y};t)q({\vf x};t)~\right]~,
\nonumber\\[-.2cm]
\\[-.25cm]
&&Q_a({\vf x},{\vf y};t)=F^{1/2}({\vf x}-{\vf y})~\frac12~
\left[ ~\bar q({\vf x};t)q({\vf y};t)-\bar q({\vf y};t)q({\vf x};t)~\right]~,
\nonumber
\end{eqnarray}
for the symmetric and anti-symmetric combinations of quarks.
Since the formfactor is a symmetric function with respect to an interchange of
coordinates ${\vf x}\to{\vf y}$ then interaction contribution in the scalar
sector
which is just the point of our interest can be written as
$$
{\cal V}^S_{int}=\frat{G}{2}~F({\vf x}-{\vf y})~\frac12~
\left[ ~\bar q({\vf x};t)q({\vf y};t)~\bar q({\vf y};t)q({\vf x};t)+
\bar q({\vf y};t)q({\vf x};t)~\bar q({\vf x};t)q({\vf
y};t)~\right]=\frat{G}{2}~(Q_s^2-Q_a^2)~.
$$
It is easy to see that now the standard procedure of bosonization may be
realized with the Gaussian integration which concerns auxiliary meson fields
$\sigma_s({\vf x},{\vf y};t)$ and
$\sigma_a({\vf x},{\vf y};t)$.
Indeed, the integration is performed over the combinations including the meson
and quark fields as $\frat{\sigma_s}{(2G)^{1/2}}-
\left(\frat{G}{2}\right)^{1/2}Q_s$,
$\frat{i\sigma_a}{(2G)^{1/2}}+\left(\frat{G}{2}\right)^{1/2}Q_a$.
Then the interaction term may be presented in the form including the meson
fields as
\begin{eqnarray}
\label{b3}
&&{\cal V}^S_{int}=\frat{G}{2}~(Q_s^2-Q_a^2)\to\nonumber\\[-.2cm]
\\[-.25cm]
&&\to-\frac{\sigma^2_s+\sigma_a^2}{2G}+\frac{\sigma_s+
i\sigma_a}{2}~F^{1/2}({\vf x}-{\vf y})~\bar q({\vf x};t)q({\vf y};t)
+\frac{\sigma_s-i\sigma_a}{2}~F^{1/2}({\vf x}-{\vf y})~\bar q({\vf y};t)q({\vf
x};t)~.\nonumber
\end{eqnarray}
and integrating over the quark fields we obtain the effective theory operating
with the mesons only. We do not show the detailed calculations here because they
follow the standard
procedure with one minor distinction which is a doubling of meson fields.
Because of the same
reason we present the succinct exposition of calculating the equation for
dynamical quark mass and
extracting meson correlator behaviour. We remind only that the first variation
of the effective
action allows us to determine the dynamical quark mass
\begin{equation}
\label{b4}
\frat{\sigma^{(1)}_s}{G} + \langle Q_s\rangle=0~,~~~
\frat{\sigma^{(1)}_a}{G} + i~\langle Q_a\rangle=0~,
\end{equation}
(here the first equation for symmetric field is equivalent to the Eq.
(\ref{sb13})).
As a result for the induced quark mass we have{\footnote{In the momentum
representation, respectively, we have
$M({\vf p})=G~\int \frat{d {\vf q}}{(2\pi)^3}~F({\vf p}-{\vf q})~i~{\mbox
Tr}~S({\vf q})$,
where $S({\vf q})$ is the quark Green function.}}
\begin{equation}
\label{mass}
M({\vf x})=F^{1/2}({\vf x})~\sigma^{(1)}_s({\vf x})~,~~~~
\sigma^{(1)}_s({\vf x}-{\vf y})=-
\frac{G}{2}~F^{1/2}({\vf x}-{\vf y})~\langle \sigma|\bar q({\vf x})q({\vf y})
+\bar q({\vf y})q({\vf x})|\sigma\rangle~.
\end{equation}
Certainly, we are interested in the real solutions and should consider only the
case when the average contribution of anti-symmetric quark combination becomes
trivial
$\langle Q_a\rangle=0$.

For the quadratic terms of effective meson Lagrangian in the scalar channel we
have
\begin{eqnarray}
\label{b5}
-\frat{\sigma^{2}_s+\sigma^{2}_a}{2G}
&+&\frat{1}{2}~\left(\frac{\sigma_s+i\sigma_a}{2}~\bar q({\vf x};t)~F^{1/2}({\vf
x}-{\vf y})~q({\vf y};t)
+\frac{\sigma_s-i\sigma_a}{2}~\bar q({\vf y};t)~F^{1/2}({\vf x}-{\vf y})~q({\vf
x};t) \right)\cdot\nonumber\\
&\cdot& \left(\frac{\sigma'_s+i\sigma'_a}{2}~\bar q({\vf x}';t')~F^{1/2}({\vf
x}'-{\vf y}')~q({\vf y}';t')
+\frac{\sigma'_s-i\sigma'_a}{2}~\bar q({\vf y}';t')~F^{1/2}({\vf x}'-{\vf
y}')~q({\vf x}';t') \right)~,\nonumber
\end{eqnarray}
here we imply that the meson fields with the primes are dependent on the
coordinates
${\vf x}'$ and ${\vf y}'$ and $t'$. Then the pairing of quark fields with
utilizing the corresponding Green
functions leads in the momentum representation (the integrations over the
corresponding 'internal' variables are
dropped) to the equation
\begin{eqnarray}
\label{b6}
&&\sigma_\alpha({\vf p},{\vf q};p_4)~
K_{\alpha,\beta}^\sigma({\vf p},{\vf q};{\vf p}',{\vf q}';p_4)~\sigma_\beta({\vf
p}',{\vf q}';p_4)=\\
&&=\frat{\sigma_s({\vf p},{\vf q};p_4)\sigma_s(-{\vf p},-{\vf q};-p_4)+
\sigma_a({\vf p},{\vf q};p_4)\sigma_a(-{\vf p},-{\vf q};-p_4)}{2G}-
\frat12{\mbox{ Tr}}S({\vf k};k_4)S({\vf k}-{\vf p}-{\vf q};k_4-
p_4)\cdot\nonumber\\
&&
\cdot\left[\frac{\sigma_s+i\sigma_a}{2}
F^{1/2}({\vf k}-{\vf p})F^{1/2}({\vf k}+{\vf q}')\frac{\sigma'_s+i\sigma'_a}{2}+
\frac{\sigma_s+i\sigma_a}{2}
F^{1/2}({\vf k}-{\vf p})F^{1/2}({\vf k}+{\vf p}')\frac{\sigma'_s-
i\sigma'_a}{2}+\right.\nonumber\\
&& 
\left.+\frac{\sigma_s-i\sigma_a}{2}
F^{1/2}({\vf k}-{\vf q})F^{1/2}({\vf k}+{\vf q}')\frac{\sigma'_s+i\sigma'_a}{2}
+\frac{\sigma_s-i\sigma_a}{2}
F^{1/2}({\vf k}-{\vf q})F^{1/2}({\vf k}+{\vf p}')\frac{\sigma'_s-
i\sigma'_a}{2}\right],\nonumber
\end{eqnarray}
where the following notations are introduced
$\sigma_{a,s}=\sigma_{a,s}({\vf p},{\vf q};p_4)$,
$\sigma'_{a,s}=\sigma_{a,s}({\vf p}',{\vf q}';-p_4)$; ${\vf q}'=-{\vf p}-{\vf
q}-{\vf p}'$,
$\alpha,\beta=s,a$.

Apparently, there is no special need to investigate the meson correlation
functions in so general form. We can obtain quite enough information on the
solutions analyzing some
particular cases. First we consider the conditions when the formfactors becomes
identical
$F^{1/2}({\vf k}-{\vf q})=F^{1/2}({\vf k}-{\vf p}-{\vf q}-{\vf p}')$,
$F^{1/2}({\vf k}-{\vf q})=F^{1/2}({\vf k}+{\vf p}')$.
These allow to conclude that the momenta of quarks coincide ${\vf p}={\vf q}$,
i.e. there is no a relative motion of quarks in such a situation. Then it is
easy to understand that the contribution of antisymmetric fields becomes
degenerate in this
configuration and the remaining symmetric part corresponds explicitly to the
standard bosonization procedure.
Finally, we have for the meson correlators in scalar and pseudoscalar channels
$$
K^{\sigma,\pi}=-\frat{1}{2G}+2N_c\int \frat{dk}{(2\pi)^4}~
F({\vf k}-{\vf p})~\frat{k_4(k_4-p_4)+{\vf k}({\vf k}-2{\vf p})\mp(m+M({\vf
k}))(m+M({\vf k}-2{\vf p}))}
{[k^2+(m+M({\vf k}))^2][(k_4-p_4)^2+({\vf k}-2{\vf p})^2+(m+M({\vf k}-2{\vf
p}))^2]},$$
$k^2=k_4^2+{\vf k}^2$.
In particular, for the Keldysh model it reads
\begin{equation}
\label{b8}
K^{\sigma,\pi}=-\frat{1}{2G}+2N_c\int \frat{dk_4}{2\pi}~
\frat{k_4~(k_4-p_4)-{\vf p}^2\mp(m+M({\vf p}))^2}
{[k_4^2+E^2({\vf p})][(k_4-p_4)^2+E^2({\vf p})]}~,
\end{equation}
where the notation $E^2({\vf p})={\vf p}^2+(m+M({\vf p}))^2$ for the quark
energy is used. To simplify the presentation of formulae we omit the energy
$E$ dependence on
the momentum. Due to the fact that only the integration over $k_4$ is essential
we say
about the one-dimensional model for mesons in this paper.

The denominator of the $\pi$-meson in Eq. (\ref{b8}) can be written in more
convenient form as
\begin{equation}
\label{b9}
\frat{1}
{[k_4^2+E^2][(k_4-p_4)^2+E^2]}=\frat{1}{\left[2~k_4~(k_4-
p_4)+p^2_4+2~E^2\right]}
\left[\frat{1}{k_4^2+E^2}+ \frat{1}{(k_4-p_4)^2+E^2}\right]~.
\end{equation}
We see the correlation function of the $\pi$-meson in Eq. (\ref{b8}) is
expressed by three integrals
$$K^\pi=-\frat{1}{2G}+2N_c\int \frat{dk_4}{2\pi}~\frat{1}{2}
\left(\frat{1}{k_4^2+E^2}+\frat{1}{(k_4-p_4)^2+E^2}\right)
-N_c\int \frat{dk_4}{2\pi}~\frat{p_4^2+4~p^2}
{[k_4^2+E^2][(k_4-p_4)^2+E^2)]}.$$
Calculating them we have finally the following result

\vspace{0.5cm}
\parbox[b]{7.in}{$ K^\pi=
 -\frat{1}{2G}+\frat{N_c}{2E}+N_c~
\left\{\begin{array}{l}
-\frat{p_4^2+4 p^2}{2 ~p_4~ E~ (p_4+2iE)}~,~~~~~Im ~p_4>i E~, \\
\frat{1}{2E}-\frat{p_4^2+4 p^2}{ E~(p_4^2+4E^2)}~,~~~~|Im~ p_4|<E~,\\
-\frat{p_4^2+4 p^2}{2 ~p_4~ E ~(p_4-2iE)}~,~~~~~Im ~p_4<-i E~,
\end{array} \right.
$}
\vspace{0.3cm}

\noindent
where $p=|{\vf p}|$. In the Euclidean domain we have for the real values of
energy $p_4$ that
\begin{equation}
\label{b11}
K^\pi=-\frat{N_c}{E~(p_4^2+4E^2)}~\left(\frat{E}{\widetilde G}~p_4^2+
\frat{E-\widetilde G}{\widetilde G}~4E^2+4 p^2\right)~,
\end{equation}
where $\widetilde G=2N_c G$. Then we find that the meson correlation function
resembles a screening factor. In order to investigate the pseudo-euclidean
situation we continue the
$p_4$ variable to the imaginary axis. Introducing the notation $p_4= i P_0$ we
have

\vspace{0.5cm}
\parbox[b]{7.in}{$ K^\pi=
N_c \left\{\begin{array}{l}
\frat{1}{E~(P_0^2-4E^2)}~\left(-\frat{E}{\widetilde G}~P_0^2+
\frat{E-\widetilde G}{\widetilde G}~4E^2+4 p^2\right)~,
~~~~~~~P_0<E, \\
\frat{1}{P_0~E~(P_0+2E)}~\left(-\frat{E}{\widetilde G}~P_0^2+
\frat{\widetilde G-2E}{\widetilde G}~E~P_0+2 p^2\right)~,
~~P_0>E,
\end{array} \right.
$}
\vspace{0.3cm}

\noindent
Comparing this expression at $P_0<E$ with Eq. (\ref{b11}) we make certain that
the transition from the Euclidean variables $p_4$ to pseudo-euclidean ones do
not change its form.
The branch $Im ~p_4<-i E$ is not be considered because of the symmetry reason.

Now we continue with searching the $\pi$-meson dispersion law which is defined
by
the zeros of correlation function $K^\pi=0$. The results for scalar and pseudo-
scalar mesons are
presented in Fig. 1 and for the vector and axial vector mesons are shown in Fig.
2. For the branch $P_0<E$ the
dispersion can be received from the following equation
\begin{equation}
\label{b12}
P_0^2=4~(E-\widetilde G)~E+\frat{\widetilde G}{E}~(2~p)^2~.
\end{equation}
It can be obtained from Eq. (\ref{sb14}) for the quark energy
\begin{equation}
\label{eq}
E=\widetilde G~\frat{m+M}{M}~.
\end{equation}
In particular, for zero quark momentum we have $M(0)=\widetilde G$ for induced
quark mass and for the $\pi$-meson energy we receive
\begin{equation}
\label{b13}
P_0^2=4~(E-\widetilde G)~E=4~m~(m+\widetilde G)~,
\end{equation}
which means that in the chiral limit ($m\to 0$) the Goldstone theorem is valid
(see. Fig. 1, curve 1). Tuning the model parameters as was proposed in Ref.
\cite{MZ} we get the
following parameters for the NJL model \cite{vmk} $m=(m_u+m_d)/2=5.$ MeV,
$\widetilde G=286$ MeV. For the
$\pi$-meson energy we have at that $P_0\approx 76$ MeV for the zero quark
momentum but the quark energy at low momenta is $E \approx m+M \approx 286$ MeV.
When $P_0>E$ the interesting
branch looks like $P_0=N_c G-E+\left[(E-N_c G)^2+\frat{4N_c G ~
p^2}{E}\right]^{1/2}$ but the
analysis shows this branch does not satisfy the constraint $P_0>E$.

Turning now to the scalar channel we present the integrand in the convenient
form as
(\ref{b8})
\begin{equation}
\label{rep2}
\frat{k_4(k_4-p_4)-E^2}
{[(k_4^2+E^2][k_4-p_4)^2+E^2]}=
\frat12~\left[\frat{1}{k_4^2+E^2}+ \frat{1}{(k_4-p_4)^2+E^2}\right]-
\frat{1}{2}\frat{p_4^2+4E^2}{[k_4^2+E^2][(k_4-p_4)^2+E^2]}~.
\end{equation}
and for the $\sigma$-meson correlation function receive
$$
K^\sigma=-\frat{1}{2G}+2N_c\int
\frat{dk_4}{2\pi}~\frat{1}{2}\left(\frat{1}{k_4^2+E^2}+
\frat{1}{(k_4-p_4)^2+E^2}\right)-N_c\int \frat{dk_4}{2\pi}~
\frat{p_4^2+4E^2}{[k_4^2+E^2][(k_4-p_4)^2+E^2]}.$$
Calculating the integrals we come to the following result

\vspace{0.5cm}
\parbox[b]{7.in}{$ K^\sigma=
 -\frat{1}{2G}+\frat{N_c}{2E}+N_c~
\left\{\begin{array}{l}
-\frat{p_4-2iE}{2 ~p_4~ E}~,~~~~~Im ~p_4>i E, \\
-\frat{1}{2E}~,~~~~~~~~~~~~|Im~ p_4|<E, \\
-\frat{p_4+2iE}{2 ~p_4~ E }~,~~~~~Im ~p_4<-i E.
\end{array} \right.
$}
\vspace{0.3cm}

\noindent
In the pseudo-euclidean regime for the branch $P_0<E$ the $\sigma$-meson
correlation function is degenerated $K^\sigma=-\frat{1}{2 G}$ and for $P_0>E$ we
have
$K^\sigma=\frat{-P_0+\widetilde G}{2G P_0}$.
Comparing to the quark energy of Eq. (\ref{eq}) we conclude that at
$P_0>E$ there are not the interesting zeros in $K^\sigma$.

We have considered the configuration when the relative momentum of quark
and anti-quark equals zero and below we address the quark and
anti-quark system with zero total momentum
 ${\vf p}+{\vf q}={\vf 0}$, see. Eq. (\ref{b6}).
For the outgoing quark momenta two configurations are possible:
a) ${\vf p}'={\vf p}$ and b) ${\vf p}'=-{\vf p}$.
In the a)-situation we obtain for the correlation functions in scalar and
pseudo-scalar channels
\begin{eqnarray}
\label{b18}
&&\pi_\alpha~K^\pi_{\alpha,\beta}~\pi_\beta=
-\frat12~\frat{\pi_s^2+\pi_a^2}{2G}+
2N_c~\frat{\pi_s^2-\pi_a^2}{2}~\int \frat{d k_4}{2\pi}
\frat{k_4(k_4-p_4)+p^2+(m+M)^2}{[k_4^2+E^2][(k_4-p_4)^2+E^2]}~,\nonumber\\
&&\sigma_\alpha~K^\sigma_{\alpha,\beta}~\sigma_\beta=
-\frat12~\frat{\sigma_s^2+\sigma_a^2}{2G}+
2N_c~\frat{\sigma_s^2-\sigma_a^2}{2}~\int \frat{d k_4}{2\pi}
\frat{k_4(k_4-p_4)+p^2-(m+M)^2}{[k_4^2+E^2][(k_4-p_4)^2+E^2]}~,\nonumber
\end{eqnarray}
where $\alpha,\beta=s,a$. Then for the $\pi$-meson we have

\vspace{0.5cm} \parbox[b]{3.6in}{$
K_s^{\pi}=
 \left \{ \begin{array}{l}
\frat{-P_0 +\widetilde G-2E}{2 G(P_0+2E)}~,~~~~~~P_0>E \\
\frat{P^2_4 -4E~(E-\widetilde G)}{2 G(4E^2-P_0^2)}~,~~P_0<E
\end{array} \right.
$}
\parbox[b]{3.6in}{$
K_a^{\pi}=
 \left \{ \begin{array}{l}
\frat{-P_0-\widetilde G-2E}{2G(P_0+2E)},~~~~~P_0>E\\
\frat{P^2_4 -4E~(E+\widetilde G)}{2 G(4E^2-P_0^2)},~P_0<E
\end{array} \right.
$}
\vspace{0.3cm}

\noindent
The dispersion law for the $K_s^\pi$ correlation function at $P_0<E$ is
extracted from the following equation
\begin{equation}
\label{dsp}
P_0^2=4~(E-\widetilde G)~E~,
\end{equation}
which is in a full agreement with Eq. (\ref{b12}) if the total momentum of quark
anti-quark pair is taken to develop  value $2 p\to 0$ (see. Fig. 1, curve 2).
For the $K_a^\pi$ correlation function there is no solution meeting the
constraint $P_0<E$ and there is no any solution for the $K_s^\pi$ correlation
function at $P_0>E$ as
well as for the branch $K_a^\pi$.

In the scalar channel one can obtain

\vspace{0.5cm} \parbox[b]{3.4in}{$
K_s^{\sigma}=
 \left \{ \begin{array}{l}
\frat{-E P_0^2+E P_0(\widetilde G-2E)+2\widetilde G(m+M)^2}{2 G E P_0(P_0+2E)},
\\
\frat{E P_0^2-4E^2(E-\widetilde G)-4\widetilde G(m+M)^2}{2G E^2(4 E^2-P_0^2)},
\end{array} \right.
$}
\parbox[b]{3.4in}{$
K_a^{\sigma}=
\left \{ \begin{array}{l}
\frat{-E P_0^2-E P_0(\widetilde G+2E)-2\widetilde G(m+M)^2}{2 G E P_0(P_0+2E)},
\\
\frat{E P_0^2-4E^2(E-\widetilde G)+4 \widetilde G (m+M)^2}{2G E^2(4 E^2-P_0^2)},
\end{array} \right.
$}
\vspace{0.3cm}

\noindent
The upper ratios are written for $P_0>E$ and the lower ones for $P_0<E$.
The dispersion law for the branch $K^s_\sigma$ at $P_0<E$ is determined by the
solution of the following equation
\begin{equation}
\label{exs}
P_0^2=4~(E-\widetilde G)~E+4~\frat{\widetilde G}{E}~(m+M)^2~,
\end{equation}
see the dashed curve in Fig. 1. We did not manage to find the appropriate
solution of dispersion equation for the correlation function $K_a^\sigma$ in
this case as the condition
$P_0<E$ is invalid. Analysis of the correlation function roots at $P_0>E$ gives
the same message
that the suitable solutions are absent. We omit the discussion of the
configuration b) because the results
already given demonstrate how rich and complicated the analysis of solution
branches could be. We would like to
mention only that the presence of bound state even for the quarks with
comparatively large momenta looks
improbable.
\begin{figure*}[!tbh]
\begin{center}
\includegraphics[width=0.5\textwidth]{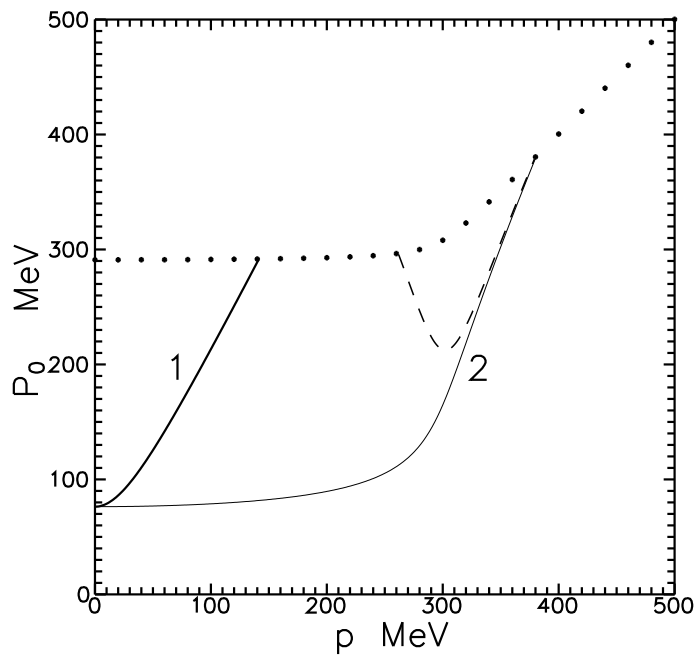}
\end{center}
\vspace{-7mm}
\caption{The $\sigma$-meson (dashed curve) and $\pi$-meson (curves $1$ and $2$)
energies in MeV as the functions of momenta. The curve $1$ corresponds to the
configuration of
zero relative quark momentum. The curve $2$ describes the quark and anti-quark
system with zero total momentum.
The dots demonstrate the quark energy $E$.}
\label{f1}
\end{figure*}

In order to calculate the pion decay constant it is necessary to calculate a
loop integral which is similar to Eq. (\ref{b8}) in which one of the vertices
responsible for the weak
interaction of the quarks does not contain the formfactor $F^{1/2}({\vf k}-{\vf
p})$
relevant for the meson fields.
Then for the Keldysh model the integral with weak singularity develops formally
the zero value what leads to the pion decay constant $f_\pi$ equal to zero.

One can consider the vector and axial-vector meson channels in the similar way
if the corresponding substitutions $\bar q q \to \bar q \gamma_\mu q$, $\sigma
\to V_\mu$,
$\bar q i\gamma_5 q \to \bar q \gamma_5\gamma_\mu q$, $\pi \to A_\mu$
are done in the relevant formulae. However, we omit those calculations here and
do use the method of the equations for vertex functions (Bethe--Salpeter
equation) to analyze the
correlation functions quantatively.

\section{Vertex functions}
It was demonstrated above that the Gaussian integration should be performed with
the symmetric and anti-symmetric combinations of auxiliary meson fields and the
anti-symmetric
fields should contain an imaginary unit factor. Clearly, the corresponding
analysis of the functional
integral saddle point in the imaginary space appears to be a rather complicated
task. Hence, it would
be highly desirable to consider some alternative possibilities of studying the
meson observables
keeping under control the quark degrees of freedom. Here we are going to use the
method based on
the Bethe--Salpeter equation. We start calculating an effective quark
interaction and sum up the set of some
special diagrams as follows
\\
\begin{picture}(50,50)(10,20)
\put(20,38){$\widetilde\Gamma=$}
\put(60,40){\line(1,1){10}}
\put(60,40){\line(1,-1){10}}
\put(60,40){\line(-1,1){10}}
\put(60,40){\line(-1,-1){10}}
\put(85,38){$+$}
\put(144,40){\line(1,1){10}}
\put(144,40){\line(1,-1){10}}
\put(120,40){\line(-1,1){10}}
\put(120,40){\line(-1,-1){10}}
\put(132,40){\circle{23}}
\put(167,38){$+$}

\put(248,40){\line(1,1){10}}
\put(248,40){\line(1,-1){10}}
\put(200,40){\line(-1,1){10}}
\put(200,40){\line(-1,-1){10}}
\put(212,40){\circle{23}}
\put(236,40){\circle{23}}
\put(267,38){$+ \cdots$}
\end{picture}
\\
\noindent
It should be taken into account from the beginning that each line of these
graphs has to be depicted as the doubled one because we consider the nonlocal
meson fields.
The first diagram of
this set describes the initial interaction
$\Gamma_0({\vf p},{\vf q})=(2\pi)^3 G~ \Gamma~F({\vf p}-{\vf q})$
where the matrix $\Gamma$ denotes the interaction channel $1$, $i \gamma_5$,
$\gamma_\mu$,
$\gamma_5\gamma_\mu$. If we single out the combination
$\bar q_{\alpha}({\vf x};t)~\widetilde \Gamma_{\alpha\beta}({\vf x},{\vf
y};t)~q_{\beta}({\vf y};t)$
then the following equation can be calculated for the series sum
\begin{eqnarray}
\label{v1}
&&-(2\pi)^3 G~ \Gamma~F({\vf p}-{\vf q})+\widetilde\Gamma(-{\vf p},-{\vf
q};r_4)=
\nonumber\\[-.2cm]
\\[-.25cm]
&&=-\int \frat{dk_4 d{\vf k}}{2\pi} G~\Gamma ~F(-{\vf k}-{\vf q})~
{\mbox {Tr}}~ S({\vf k};k_4)~\Gamma~ S({\vf k}-{\vf p}+{\vf q};k_4-r_4)
~\widetilde\Gamma({\vf k}-{\vf p}+{\vf q},{\vf k};k_4-r_4),\nonumber
\end{eqnarray}
where $r_4=p_4-q_4$ and $p_4$, $q_4$ are the corresponding components of quark
and anti-quark momenta. We search for the solution for the vertex function in
the Keldysh model, for example,
in the form $\widetilde\Gamma({\vf p},{\vf q};r_4)=(2\pi)^3 ~ \Gamma~\delta({\vf
p}-{\vf q})~V({\vf p};r_4)$
assuming that for the imaginary values of $r_4$ the solution for the vertex
function possesses the pole singularity. Picking out the singular contributions
we can obtain approximately
that $-G+V=G~\Pi~V$ where the polarization operator in the Keldysh model can be
represented in the
form $\Pi=\int \frat{dk_4}{2\pi}~{\mbox {Tr}}~  S(-{\vf p};k_4)~\Gamma~S(-{\vf
p};k_4-r_4)~\Gamma$.
Therefore we have for the vertex function $V=\frat{G}{1-G~\Pi}$ and its
denominator zeros determine the pole positions.
Discussing the bosonization above we have already calculated the similar
polarization operators. Omitting the intermediate calculaions we show here the
results for pseudo-scalar
and scalar channels at $|r_4|<E$ as
\begin{eqnarray}
\label{pol}
&&\frat{1-G~\Pi^{{\mbox {\tiny PS}}}}{\widetilde G}=
\frat{4E~(E-\widetilde G)-R_0^2}{\widetilde G~(4E^2-R_0^2)}~,\nonumber\\
&&\frat{1-G~\Pi^{{\mbox {\tiny S}}}}{\widetilde G}=
\frat{(4E^2-R_0^2)~E+4[(m+M)^2-E^2]~\widetilde G}{\widetilde G~E~(4E^2-
R_0^2)}~,\nonumber
\end{eqnarray}
where $R_0=P_0-Q_0$, $P_0=ip_4$, $Q_0=iq_4$.
In the pseudo-scalar channel the $\pi$-meson dispersion law coincides explicitly
with Eq. (\ref{dsp}) and for the $\sigma$-meson we have Eq. (\ref{exs}).

For vector and axial-vector channels in the Keldysh model we have
\begin{eqnarray}
\label{vav}
&&\Pi^{{\mbox {\tiny V,A}}}_{44}=-4N_c~\int \frat{dk_4}{2\pi}~
\frat{k_4(k_4-r_4)- p^2\mp (m+M)^2}
{[k_4^2+E^2][(k_4-r_4)^2+E^2]}~,\nonumber\\
&&\Pi^{{\mbox {\tiny V,A}}}_{4i}=4N_c~\int \frat{dk_4}{2\pi}~
\frat{(2 k_4-r_4)~p_i}{[k_4^2+E^2][(k_4-r_4)^2+E^2]}~,\\
&&\Pi^{{\mbox {\tiny V,A}}}_{ij}=4N_c~\int \frat{dk_4}{2\pi}~
\frat{k_4(k_4-r_4)~\delta_{ij}-2~p_i p_j+\delta_{ij}~ p^2\pm
\delta_{ij}~(m+M)^2}
{[k_4^2+E^2][(k_4-r_4)^2+E^2]}~,\nonumber
\end{eqnarray}
$\Pi^{{\mbox {\tiny V,A}}}_{4i}=\Pi^{{\mbox {\tiny V,A}}}_{i4}$.
It is easy to see that these results coincide with the corresponding meson
correlation function obtained above. Calculating the integrals in
vector and axial-vector channels for $Im~ r_4>iE$ we have:
\begin{eqnarray}
\label{v10}
&&\Pi^{{\mbox {\tiny V}}}_{44}=-2 N_c ~\frat{i}{r_4}~,~~~~~~~~~~~~~~~~~~~
\Pi^{{\mbox {\tiny A}}}_{44}=-\frat{2 N_c }{E}~\frat{i~ r_4 E-2
p^2}{r_4(r_4+2iE)}~,
\nonumber\\
&&\Pi^{{\mbox {\tiny V}}}_{4i}=-2 N_c~\frat{ p_i}{r_4~E}~,~~~~~~~~~~~~~~~~
\Pi^{{\mbox {\tiny A}}}_{4i}=\Pi^{{\mbox {\tiny V}}}_{4i}~,
\\
&&\Pi^{{\mbox {\tiny V}}}_{ij}=2 N_c ~
\frat{i~ r_4 E~\delta_{ij}-2 p_i p_j}{r_4(r_4+2i E)E}~,~~~
\Pi^{{\mbox {\tiny A}}}_{ij}=\frat{2 N_c }{E}~
\frat{[i~ r_4E-2(m+M)^2]\delta_{ij}-2 p_i p_j}{r_4(r_4+2i E)}~.
\nonumber
\end{eqnarray}

At $|r_4|<E$ we obtain
\begin{eqnarray}
\label{v11}
&&\Pi^{{\mbox {\tiny V}}}_{44}=0~,~~~~~~~~~~~~~~~~~~~~~~~~~~~~
\Pi^{{\mbox {\tiny A}}}_{44}=-\frat{8 N_c }{E}~\frat{(m+M)^2}{r_4^2+4 E^2}~,
\nonumber\\
&&\Pi^{{\mbox {\tiny V}}}_{4i}=0~,~~~~~~~~~~~~~~~~~~~~~~~~~~~~
\Pi^{{\mbox {\tiny A}}}_{4i}=\Pi^{{\mbox {\tiny V}}}_{4i}~, \\
&&\Pi^{{\mbox {\tiny V}}}_{ij}=8 N_c~\frat{E^2~\delta_{ij}-p_i p_j}{E~(r_4^2+4
E^2)}~,~~
\Pi^{{\mbox {\tiny A}}}_{ij}=\frat{8 N_c }{E}~\frat{p^2\delta_{ij}- p_i p_j}
{r_4^2+4 E^2}~.\nonumber
\end{eqnarray}

\begin{figure*}[!tbh]
\begin{center}
\includegraphics[width=0.5\textwidth]{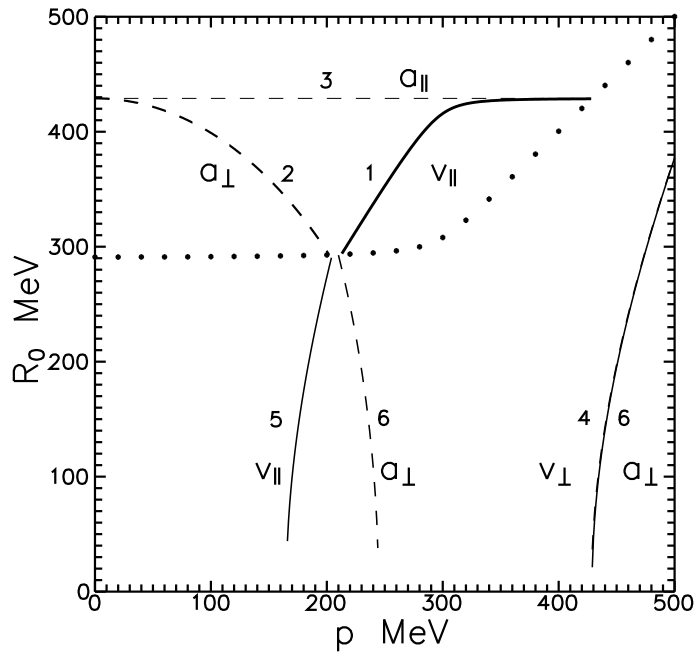}
\end{center}
\vspace{-7mm}
\caption{The relative energies $R_0$ of vector (solid curves) and axial-vector
(dashed curves)  mesons (in MeV), as the functions of momenta (in MeV). The
notations are explained in
the text and the dots denote the quark energy $E$.}
\label{f2}
\end{figure*}

Now we diagonalize the correlation functions using the fact that corresponding
quadratic forms determine simply the Lagrangian of free vector and axial-vector
mesons
\begin{equation}
\label{form}
K^{\mbox {\tiny V,A}}=\widetilde C^{\mbox {\tiny V, A}}_{44}~V_4^2+2~\widetilde
C^{\mbox {\tiny V, A}}_{4i}~V_4\widetilde V_i+
\widetilde V_i~ \widetilde C^{\mbox {\tiny V, A}}_{ij}~\widetilde V_j~,
\end{equation}
where we imply the summation over the indices which are repeated.
It is valid by definition that
$\widetilde C^{\mbox {\tiny V,A}}_{\mu\nu}=\delta_{\mu\nu}-
G_{\mbox {\tiny V}}~\Pi^{\mbox {\tiny V, A}}_{\mu\nu}$.
If we redefine the space components of (axial-)vector fields by substituting
$\widetilde V_i=V_i+\alpha~ p_i V_4$, where
$\alpha^{\mbox {\tiny V, A}}=-\frat{\widetilde C^{\mbox {\tiny V, A}}_{4i}~p_i}
{p_i ~\widetilde C^{\mbox {\tiny V, A}}_{ij}~ p_j}$,
and exclude the mixed components $V_4 V_i$ from quadratic form (\ref{form}) we
get that the fourth component of vector field enters the quadratic form with
coefficient
$C^{\mbox {\tiny V, A}}_{44}=\widetilde C^{\mbox {\tiny V, A}}_{44}-
\frat{(\widetilde C^{\mbox {\tiny V, A}}_{4i}~p_i)^2}{p_i~ \widetilde C^{\mbox
{\tiny V, A}}_{ij}~ p_j}$.
The components of the tensor $C^{\mbox {\tiny V, A}}_{ij}$ remain unchanged.
The numerical analysis demonstrates the acceptable solution for the dispersion
of the fourth component exists only for the axial-vector field at $R_0>E$.
However, we do not
discuss this solution in the present paper. The spatial components of vector
fields are searched as having
two different forms for the transversal
$V_i=\left(\delta_{ij}-\frat{p_i p_j}{ p^2}\right)~v^{\bot}_j$ and longitudinal
$V_i=p_i ~v^{\|}$ components. First we consider the case of $R_0>E$.
The dispersion law for the transversal component of vector field $v^{\bot}$ has
the form $R_0=\widetilde G_{\mbox {\tiny V}}-2 E$,
where $\widetilde G_{\mbox {\tiny V}}=2 N_c G_{\mbox {\tiny V}}$.
Therefore the nontrivial solution is possible when the condition $\widetilde
G_{\mbox {\tiny V}}>3 E$ is satisfied.
For clarity we take the constant as
$\widetilde G_{\mbox {\tiny V}}=1.5 \widetilde G$ which corresponds at the
low quark momentum to the value $\widetilde G_{\mbox {\tiny V}}\approx 1.5 E$
and therefore for the fitting parameter set selected this solution branch does
not manifest itself.
The dispersion of longitudinal component $v^{\|}$ in this case is defined as
$R_0=N_c G_{\mbox {\tiny V}}- E+\left[(E-N_c G_{\mbox {\tiny V}})^2+
\frat{4 N_c G_{\mbox {\tiny V}} p^2}{E}\right]^{1/2}$.
The branches of meson observables in the vector and axial-vector channels are
depicted in Fig. \ref{f2}.
The branch discussed above is shown by the solid curve and denoted by $1$.
For the transversal component of axial-vector meson $a^{\bot}$ we obtain
$R_0=N_c G_{\mbox {\tiny V}}- E+\left[(E-N_c G_{\mbox {\tiny V}})^2+
\frat{4 N_c G_{\mbox {\tiny V}} (m+M)^2}{E}\right]^{1/2}$.
This curve is depicted in Fig. \ref{f2} by dashed line and marked by $2$.
The dispersion of the longitudinal component $a^{\|}$ is calculated from
$R_0=\widetilde G_{\mbox {\tiny V}}$. This branch appears when the condition
$\widetilde G_{\mbox {\tiny V}}>E$ is
satisfied. In Fig. \ref{f2} it corresponds to the dashed straight line $3$. We
see
at low momentum the longitudinal  $a^{\|}$ and transversal $a^{\bot}$ components
practically coincides.

Now we address the situation of $R_0<E$.
The dispersion of the transversal component of the vector meson $v^{\bot}$ is
defined by $R_0^2=4E~(E-\widetilde G_{\mbox {\tiny V}})$.
This curve is shown in Fig. \ref{f2} by solid line $4$.
For the longitudinal components $v^{\|}$ we have
$R_0^2=4~\left(E^2-\widetilde G_{\mbox {\tiny V}}\frat{(m+M)^2}{E}\right)$,
and the solid curve $5$ shows its behaviour.
For the transversal component of the axial-vector meson $a^{\bot}$ we obtain
$R_0^2=4E^2-8N_c~G_{\mbox {\tiny V}}~\frat{ p^2}{E}$.
The corresponding curve is presented as the dashed line $6$. Its right-hand
component is practically degenerated with the curve $4$ because the induced
quark mass goes to zero at
large momenta and the restoration of chiral symmetry takes place.
The longitudinal component of axial-vector field $a^{\|}$ is degenerated.

The $\pi$-meson energy in the NJL model for the tuning parameter values
considered coincides with experimental data and looks like $E^{{\mbox {\tiny
NJL}}}_\pi=140$ MeV.
In Ref. \cite{MZ} it was supposed that for the Keldysh model the dynamical quark
mass in the low momentum region is equal to the dynamical quark mass of the NJL
model. This assumption
has led to the almost identical quasiparticles for both models. But the $\pi$-
meson energy turned out rather
underestimated  $E^{{\mbox {\tiny K}}}_\pi=76$ MeV although the relative scale
of mesons for different channels was maintained. The axial-vector meson had
gotten heavier than the vector meson and the meson of pseudoscalar channel was
the lightest one, indeed.

Calculating the correlation functions here we did not exploit the convenient
trick of shifting the integration variable to make the integrals symmetric and
were keeping the
integration contour fixed. The one-dimensional character of the model provides
us with the
obvious possibility to study the dependence of the correlation functions on the
integration
contour shape as well. In particular, it is interesting to trace the turn of
integration
contour to the imaginary axis. Then two quark poles could also be treated as the
"Wigner"
phase and the continuation of correlation
functions will be different from what has been done in this note.

\section{Correlation functions in the Minkowski space}
The model developed allows us to study easily the meson correlation functions in
the Minkowski space
as well and to compare them to what we obtained above. In fact, the task is
technically related to
computing the following integral (dependent on the Euclidean variables) within
the fixed contour
\begin{eqnarray}
&&I=\int^\infty_{-\infty}~\frat{dq}{2\pi}~\frat{1}{(q^2+E^2)[(q-
p)^2+E^2]}=\nonumber\\
&&=\int^\infty_{-\infty}~\frat{dq}{2\pi}~\frat{1}{(2iE)^2}~
\left[\frat{2iE}{p(p-2iE)}\frat{1}{q-iE}+
\frat{2iE}{p(p+2iE)}\frat{1}{q-p-iE}+
\right.\nonumber\\
&&~~~~~~~~~~~~~~~~~~~~~~~~~
+\left.\frat{-2iE}{p(p+2iE)}\frat{1}{q+iE}+
\frat{-2iE}{p(p-2iE)}\frat{1}{q-p+iE}\right]~\nonumber
\end{eqnarray}
which is the sum of four pole terms:
$i E$ designated as $e_1$, $p+iE$ indicated as $e_2$,
$-i E$ marked as $m_1$ and $p-iE$ signed as $m_2$. It is implied the
parameter $p$ may take the
complex values as well. Then the calculation of this integral leads to\\
\vspace{0.5cm}
\parbox[b]{7.in}{$ I=
 \frat{-i}{4 E^2}~
\left\{\begin{array}{l}
\frat{2 i E}{p(p+2iE)}~,~~~~~~Im~ p>i E~, \\
\frat{4iE}{p^2+4 E^2}~,~~~~~~~~|Im~ p|<E~,\\
\frat{2i E}{p(p-2iE)}~,~~~~~~~Im~p<-i E~.
\end{array} \right.
$}
\vspace{0.3cm}

\noindent
Our concern here is the particular situation when the parameter $p$ is pure
imaginary $p=iP$. Now we should calculate the similar integral with another
fixed contour which
corresponds to the Minkowski space (turned to 90 degrees regarding the Euclidean
integration)
contour
\begin{eqnarray}
&&J=\int^\infty_{-\infty}~\frat{dq}{2\pi}~\frat{i}{(q^2-E^2+i\varepsilon)
[(q-p)^2-E^2+i\varepsilon]}=\nonumber\\
&&=\int^\infty_{-\infty}\frat{dq}{2\pi}\frat{i}{[2(E-i\varepsilon)]^2}
\left[\frat{2(E-i\varepsilon)}{p[p-2(E-i\varepsilon)]}\frat{1}{q-
E+i\varepsilon}+
\frat{2(E-i\varepsilon)}{p[p+2(E-i\varepsilon)]}\frat{1}{q-p-E+i\varepsilon}+
\right.\nonumber\\
&&~~~~~~~~~~~~~~~~~~~~~~~~~~~~~~
+\left.\frat{-2(E-i\varepsilon)}{p[p+2(E-i\varepsilon)]}\frat{1}{q+E-
i\varepsilon}+
\frat{-2(E-i\varepsilon)}{p[p-2(E-i\varepsilon)]}\frat{1}{q-p+E-
i\varepsilon}\right]~.\nonumber
\end{eqnarray}
and its calculation gives

\vspace{0.5cm}
\parbox[b]{7.in}{$ J=
 \frat{1}{4 E^2}~
\left\{\begin{array}{l}
\frat{2 E}{p[p-2(E-i\varepsilon)]}~,~~~~~~Im~ p>i \varepsilon~, \\
\frat{4E}{p^2-4 (E-i\varepsilon)^2}~,~~~~~~~|Im~ p|<\varepsilon~,\\
\frat{2 E}{p[p+2(E-i\varepsilon)]}~,~~~~~~~Im~p<-i \varepsilon~.
\end{array} \right.
$}
\vspace{0.3cm}

\noindent
Compared to the Euclidean configuration the pole contributions to the integral
in the Minkowski space are interchanged, i.e. the poles of $e$-type become the
poles of
$m$-type. Thus, the result obtained for the Minkowski space will be valid for
the
Euclidean configuration if the contribution of the $m_2$-pole at $Im~p>iE$ is
omitted and the
contribution of $e_2$-pole at $Im~p<-iE$ is added. One should not change
anything at $|Im~p|<E$. The figure shows one of possble integration contours corresponding to the
situation $Im~p>iE$. Similarly at $Im~p<-iE$ the contour should be deformed in
order to have the
contribution of $e_2$-pole. The detailed analysis make possible to formulate the
general rule for
reproducing the proper result in the Minkowski space. The integration contour
should
be deformed in such a way to have the contributions of the $e_1$ and $e_2$-
poles only,
i.e. the integration contour looks like being squeezed in between the poles
$e_2$ and $m_2$.
\\
\begin{tabular}{ll}
&\hspace{5.cm}
\begin{picture}(100,250)(10,-30)
\put(149,50){\line(-1,0){200}}
\put(151,50){\line(1,0){50}}
\put(50,50){\line(0,1){100}}
\put(50,50){\line(0,-1){100}}
\put(60,100){$e_1$}
\put(50,100){\circle{2}}
\put(60,0){$m_1$}
\put(50,0){\circle{2}}
\put(160,190){$e_2$}
\put(150,190){\circle{2}}
\put(160,70){$m_2$}
\put(150,70){\circle{2}}
\put(149,50){\line(0,1){15}}
\put(151,50){\line(0,1){15}}
\put(150,70){\circle{9}}
\put(50,150){\line(-1,-5){2}}
\put(50,150){\line(1,-5){2}}
\put(100,50){\line(-5,1){10}}
\put(100,50){\line(-5,-1){10}}
\put(200,50){\line(-5,1){10}}
\put(200,50){\line(-5,-1){10}}
\end{picture}
\end{tabular}
\\
\\
\\
In order to calculate the integrals $I$ and $J$ in $x$-representation we rewrite
the integral $I$ (using the well-known identity and implying the analytical
continuation of all
auxiliary functions in the parameter $p$) as
\begin{eqnarray}
&&I=\int^\infty_{-\infty}\frat{dq}{2\pi}\frat{1}{(q^2+E^2)[(q-p)^2+E^2]}=
\int_0^{1}dx\int^\infty_{-\infty}\frat{dq}{2\pi}
\frat{1}{\{x(q^2+E^2)+(1-x)[(q-p)^2+E^2]\}^2}=\nonumber\\
&&=\frat{1}{4p^3}\int_{-1/2}^{1/2}d\xi
\frat{1}{\left[\frat14 +\frat{E^2}{p^2}-
\xi^2\right]^{3/2}}=\frat{1}{E(p^2+4E^2)}~,\nonumber
\end{eqnarray}
where $\xi=x-1/2$. Such an integral treatment makes transparent that the result
above corresponds to the calculation of $I$ in the fixed contour for the branch
when $|Im~p|<E$. Changing
the variables of integration as $E^2\to-E^2$ we are able to reproduce the
corresponding result
for $J$. Thus, we may conclude that calculating in the $x$-representation fully
rereproduces the result for the Minkowski space being analytically continued.
\begin{figure*}[!tbh]
\begin{center}
\includegraphics[width=0.5\textwidth]{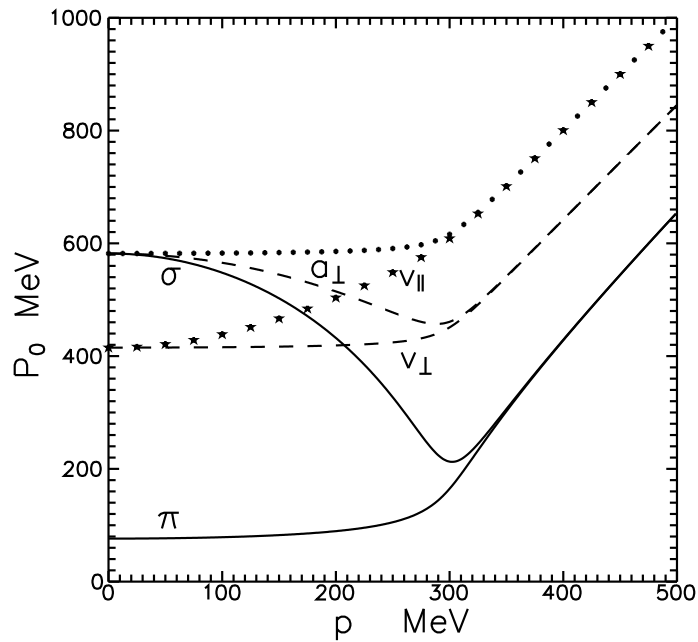}
\end{center}
\vspace{-7mm}
\caption{The meson energies obtained by using the correlation functions in the
Minkowski space as the functions of quark momentum. The double quark energy is
depicted by dots.
The dispersion law for the longitudinal component of vector meson field is
depicted by stars and
the dashed line with the symbol $v_{\bot}$ corresponds to the transverse
component. The dashed line
with the symbol $a_{\bot}$ shows the dispersion law for the transverse component
of axial vector
meson. The solid line with the symbol $\sigma$ is devoted for the scalar meson
whereas the
similar line with the symbol $\pi$ corresponds to $\pi$-meson.
}
\label{f3}
\end{figure*}

Considering the meson correlation functions in the Minkowski space we are
interested in the situation when the external parameters are real what
corresponds to the
constraint $|Im~ p|<\varepsilon$ for the $J$ integral. It is a pretty simple
task to obtain
the final results for the corresponding dispersion laws using the results of
calculations
for the Euclidean space. Below we show as an example the results for the (anti-
)quark total
momentum equal zero
\begin{eqnarray}
&&~P_\pi^2=4 E(E-\widetilde G)~,~~~~P^2_\sigma=4 E^2- ~4\frat{\widetilde
G}{E}~{\vf p}^2~,\nonumber\\
&&P_{\mbox {\tiny V}_{\bot}}^2=4 E(E-\widetilde G_{\mbox {\tiny V}})~,
~~P_{\mbox {\tiny A}_{\bot}}^2=4 E^2-4\frat{\widetilde G_{\mbox {\tiny
V}}}{E}~{\vf p}^2~,\nonumber\\
&&P_{\mbox {\tiny V}_{\|}}^2=4 E^2-4\frat{\widetilde G_{\mbox {\tiny
V}}}{E}~(m+M)^2~,\nonumber
\end{eqnarray}
the axial vector field correlator $a_{\|}$ becomes degenerate.
At $\widetilde G=\widetilde G_{\mbox {\tiny V}}$ $\pi$-meson becomes degenerate
with the vector meson and $\sigma$-meson with the axial vector meson. It is
clear the realistic
relations between the meson masses correspond to the situation when $\widetilde
G_{\mbox {\tiny
V}}<\widetilde G$. In order to give another example we take
$\widetilde G_{\mbox {\tiny V}}=\widetilde G/2$ (in
addition to the Euclidean consideration in which the constant was taken as
$\widetilde G_{\mbox {\tiny V}}=1.5~\widetilde G$.
The meson energies as quark momentum functions are shown in Fig. \ref{f3} in the
Minkowski space. Curiously, the bound states of qurk and anti-quark do exist
at any quark momentum for the present configuration.

\section{Conclusion}
In this note we demonstrate that despite the singular behaviour of mean energy
and quark condensate which was observed in Ref. \cite{MZ} the meson observables
are finite, well identified and compatible with the experimental energy scale.
The number of effective degrees of freedom which define the quasi-particle
picture in the NJL and Keldysh model are comparable. The Keldysh model being
as simple as the NJL model looks like a suitable candidate for describing the
nonequilibrium processes in the (anti-)quark ensembles. Due to the one-
dimensional
character of the Keldysh model the analytical continuation from the Euclidean
region
of meson observables to the pseudo-euclidean one is easily
performed and controled. The amazing feature of our consideration is that the
bound states are revealed  at any quark momenta in Minkowsky space.
Eventually we conclude that if the quasi-particles in the different models are
similar the meson observables are also alike.


\end{document}